\documentclass[a4paper]{jpconf}
\usepackage{graphicx}
\usepackage{amssymb}
\newcommand{\s}{\sqrt{s}}
\newcommand{\sNN}{\sqrt{s_{\scriptscriptstyle \rm NN}}}

\newcommand{\gevc}{\mathrm{GeV}/c}

\newcommand{\pPb}{\mbox{p--Pb}}

\newcommand{\Pt}{p_{\rm T}}
\newcommand{\PtD}{p_{\rm T}^{\rm D}}
\newcommand{\PtAssoc}{p_{\rm T}^{\rm assoc}}

\newcommand{\Dz}{{\rm D}^0}
\newcommand{\Ds}{{\rm D}^{*+}}
\newcommand{\Dp}{{\rm D}^+}

\newcommand{\deta}{\Delta \mathrm{\eta}}

\begin{document}
\title{Heavy-flavour production in pp collisions and correlations in pp and p-Pb collisions measured with ALICE at the LHC}

\author{Fabio Colamaria, on behalf of the ALICE Collaboration}

\address{INFN, Sezione di Bari, Via E. Orabona 4, 70125 Bari, Italy}

\ead{fabio.colamaria@ba.infn.it}

\begin{abstract}
Thanks to its excellent tracking and particle identification performance, the ALICE detector is capable of measuring D mesons at mid-rapidity via their hadronic decay channels down to very low transverse momentum.
We show an extension of the prompt $\Dz$ $\Pt$-differential cross section measurement down to zero transverse momentum, which allows us to determine the $\Pt$-integrated charm production cross section at mid-rapidity in pp collisions at \mbox{$\s$ = 7} TeV. We also present measurements of azimuthal correlations of prompt $\Dz$, $\Dp$ and $\Ds$ mesons with charged hadrons in pp collisions at \mbox{$\s$ = 7} TeV and $\pPb$ collisions at \mbox{$\sNN$ = 5.02} TeV and compare the results with expectations from models.
\end{abstract}

\section{Introduction}
\label{sec:intro}
The study of heavy-flavour production in pp collisions at LHC energies allows us to test perturbative QCD (pQCD) calculations and it provides a reference for studies in heavy-ion collisions. At low $\Pt$, heavy-flavour production is dominated by low-$x$ gluons. Measurements in this region can thus provide constraints on the pQCD calculations and gluon parton distribution functions at low-$x$ values.
An analysis technique without selections on the displaced vertex topology allowed measurements of the $\Pt$-differential prompt $\Dz$ production cross section down to $\Pt$ = 0~$\gevc$~\cite{arXivLowPt} in pp collisions at $\s$ = 7 TeV with ALICE, extending the $\Pt$ reach of previous measurements~\cite{D_pp1}. This also allowed an evaluation of the total charm production cross section with a reduced uncertainty with respect to the previous ALICE measurement~\cite{D_pp3}.
More differential measurements of charm production provided by ALICE are now also available~\cite{D_ppmult,arXivDh}. In particular, the analysis of angular correlations between heavy-flavour particles and charged particles is a tool to characterize the heavy-quark fragmentation process and is sensitive to their production mechanism. In $\pPb$ collisions, this measurement allows us to investigate possible effects due to the presence of a nucleus in the collision (cold nuclear matter effects) on the heavy-quark production and hadronisation.


\section{Prompt $\Dz$ production cross section down to $\Pt$ = 0~GeV/$c$}
\label{sec:multiplicity}
\begin{figure}[!h]
\centering
\begin{minipage}{\linewidth}
  \centering
  $\vcenter{\hbox{\includegraphics[width=.46\linewidth]{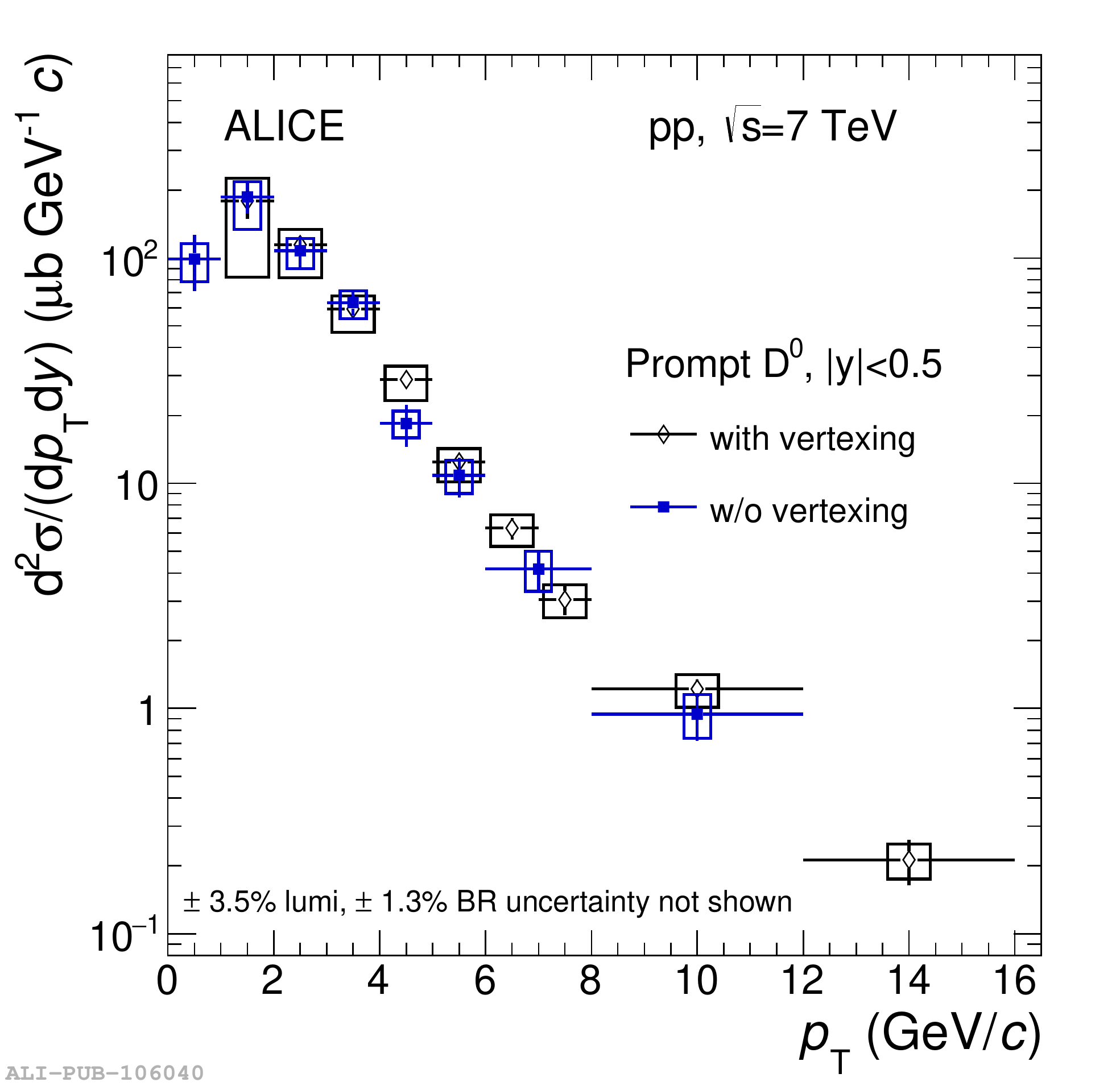}}}$
  $\vcenter{\hbox{\includegraphics[width=.46\linewidth]{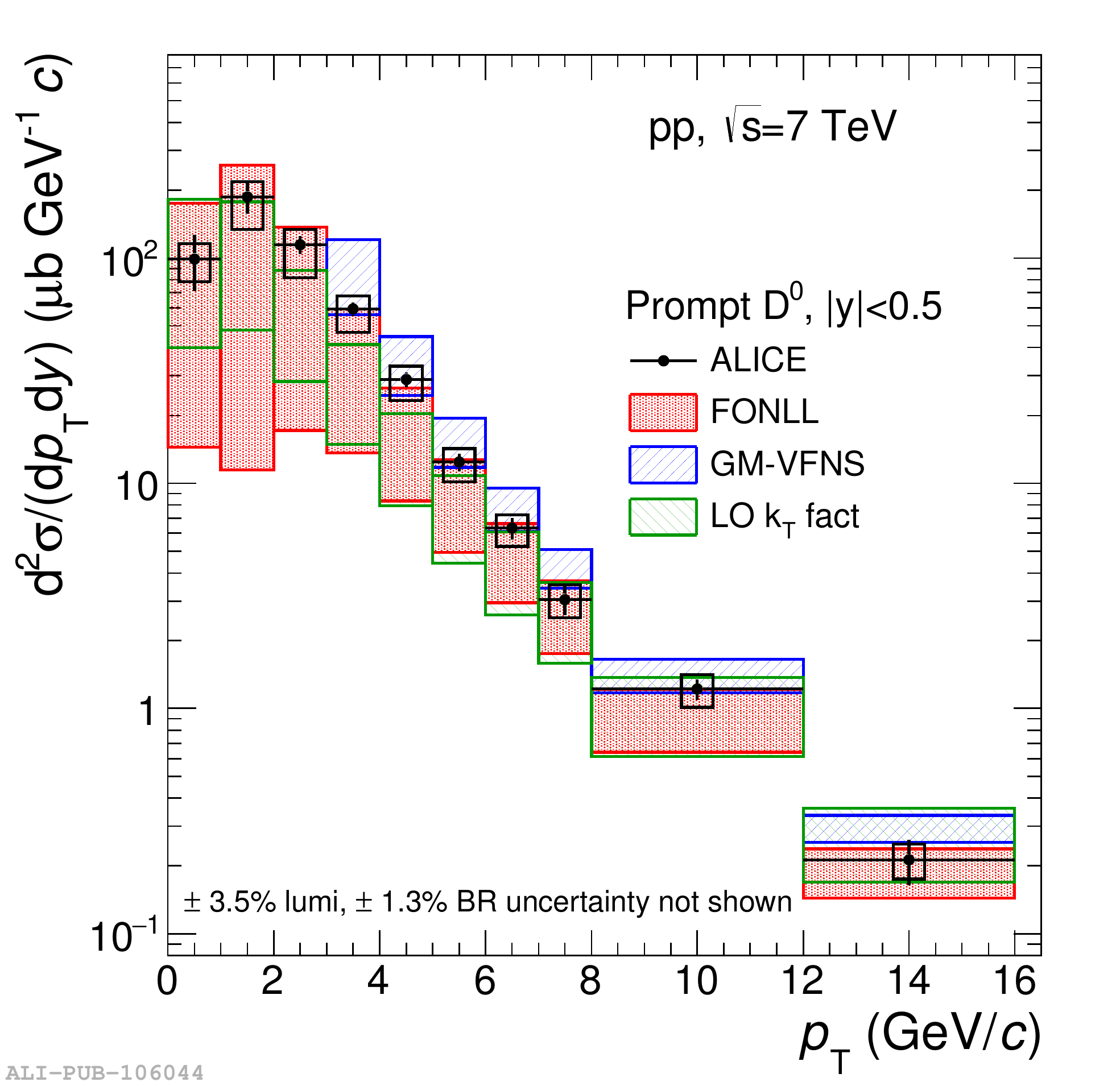}}}$
\end{minipage}

\caption{Left panel: comparison of the $\Pt$-differential production cross section for $\Dz$ mesons with $|y| < 0.5$ in pp collisions at $\s$ = 7 TeV obtained with the two analysis techniques~\cite{arXivLowPt}. Right panel: comparison of the most precise measurements of the cross section in each $\Pt$ interval with predictions by pQCD calculations~\cite{arXivLowPt}.}
\label{fig:1}
\end{figure}

In ALICE~\cite{ALICE}, D mesons are reconstructed from their hadronic decay channels (D$^0\rightarrow$ K$^-\pi^+$, D$^+\rightarrow$ K$^-\pi^+\pi^+$, D$^{*+}\rightarrow$ D$^0 \pi^+\rightarrow$ K$^-\pi^+\pi^+$, D$_{\rm s}^+\rightarrow \phi\pi^+ \rightarrow$ K$^-$K$^+\pi^+$ and their charge conjugates). The standard analysis~\cite{D_pp1} is based on (i) the selection of D-meson decay products by exploiting particle identification and track-quality cuts, and ii) a selection exploiting the \mbox{D-meson} decay length of a few hundred $\mu$m.
At very low D-meson $\Pt$, this technique is not efficient due to the small Lorentz boost of the D mesons and the degraded track impact parameter resolution. A different technique, relying only on the particle identification without any topological selection, is used to evaluate the $\Dz$ meson $\Pt$-differential production cross section down to $\Pt$ = 0~$\gevc$.
For each $\Pt$ interval, the $\Dz$-meson raw yield is extracted from an analysis of the $\Dz$ candidate invariant-mass distribution. The central value of the yield is obtained as the arithmetic average of the raw yields obtained with four different approaches used for the evaluation of the background distribution (like-sign pairs, event mixing, track rotation and side-band fit).
With this analysis technique prompt and feed-down $\Dz$ mesons are reconstructed with the same efficiency, differently from the standard approach, which favours $\Dz$ candidates from beauty-hadron decays which on average are more displaced. As a consequence, the feed-down subtraction uncertainty on the $\Pt$-differential cross section is reduced.
The left panel of Fig.~\ref{fig:1} compares the $\Pt$-differential production cross section for $\Dz$ mesons at mid-rapidity in pp collisions at $\s$ = 7 TeV obtained with the two analysis techniques~\cite{arXivLowPt}. The results are compatible within the uncertainties in the common $\Pt$ range. The analysis without secondary vertex reconstruction, however, enables the study of the $0 < \PtD < 1$ $\gevc$ interval, and provides smaller uncertainty for the $1 < \PtD < 2$ $\gevc$ interval.
In the right panel of Fig.~\ref{fig:1} the most precise measurement of the cross section is shown, exploiting for each $\Pt$ interval the measurement with the smallest total uncertainty. ALICE measurements are compared to predictions from FONLL~\cite{FONLL}, GM-VFNS~\cite{GMVFNS} and $k_{\rm t}$-factorization~\cite{kTfact} pQCD calculations, which are observed to be in agreement with data results within the theoretical and experimental uncertainties.
The resulting total $c\bar{c}$ production cross section in pp collisions at $\s$ = 7 TeV is $\sigma^{\rm tot}_{c\bar{c}} = 7.96 \pm 0.65\mathrm{(stat.)} ^{+0.87}_{-1.57}\mathrm{(syst.)}^{+2.34}_{-0.35}\mathrm{(extr.)}\pm 0.28\mathrm{(lumi.)}\pm 0.10\mathrm{(BR)}\pm 0.03\mathrm{(FF)}$ mb, with reduced systematic and extrapolation uncertainties with respect to the previous measurement~\cite{D_pp3}.

\section{D meson-charged particle angular correlations}
\label{sec:correlations}
\begin{figure}[!p]
\centering
    \includegraphics[width=0.77\textwidth]{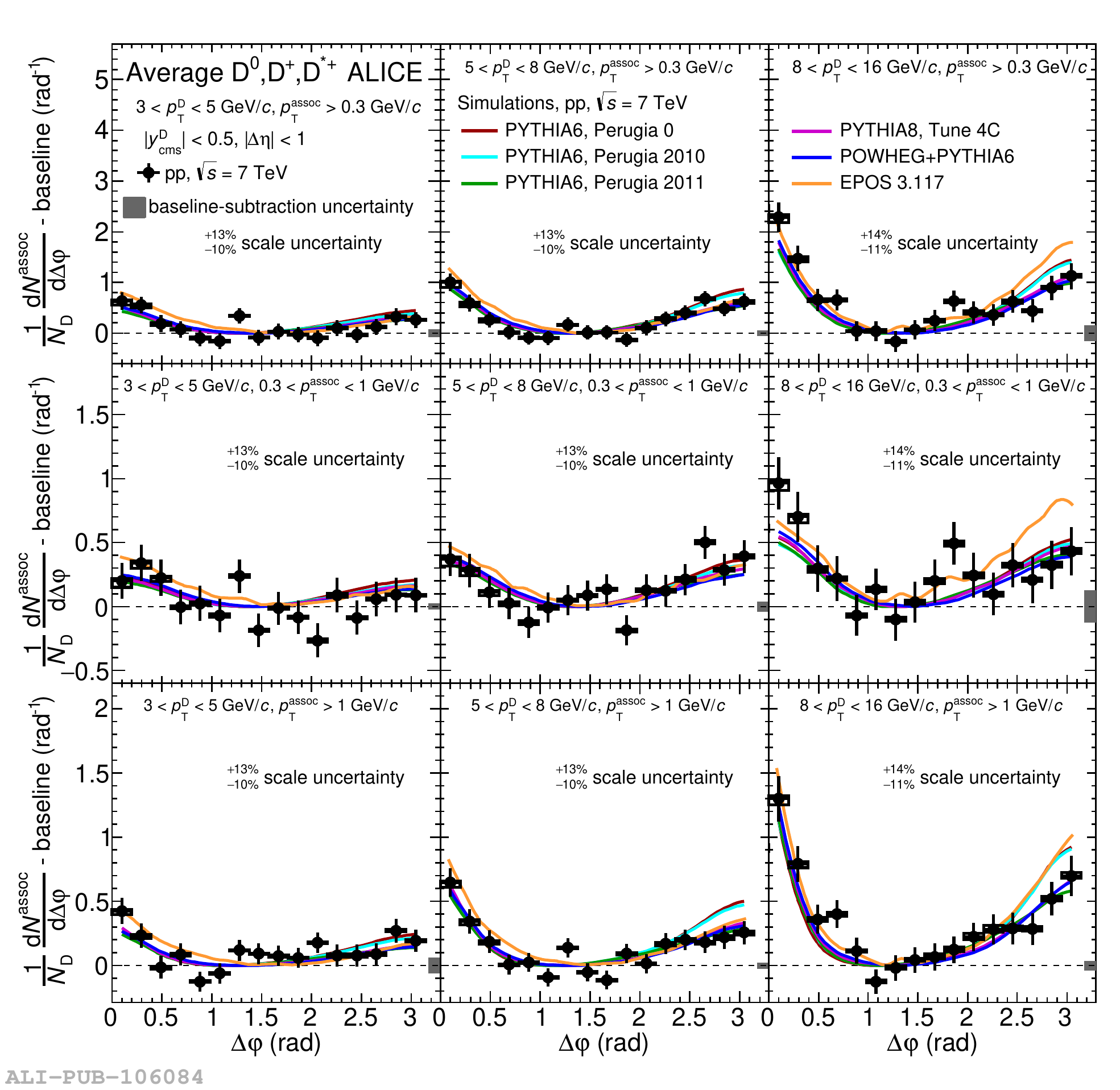}
    \vspace{-0.5em}
    \caption{Comparison of D meson-charged particle azimuthal correlation distributions in pp collisions with predictions from Monte Carlo simulations, for different D-meson and associated particle $\Pt$ ranges~\cite{arXivDh}.}
    \label{fig:2}
\end{figure}

\begin{figure}[!p]
\centering
    \includegraphics[width=0.77\textwidth]{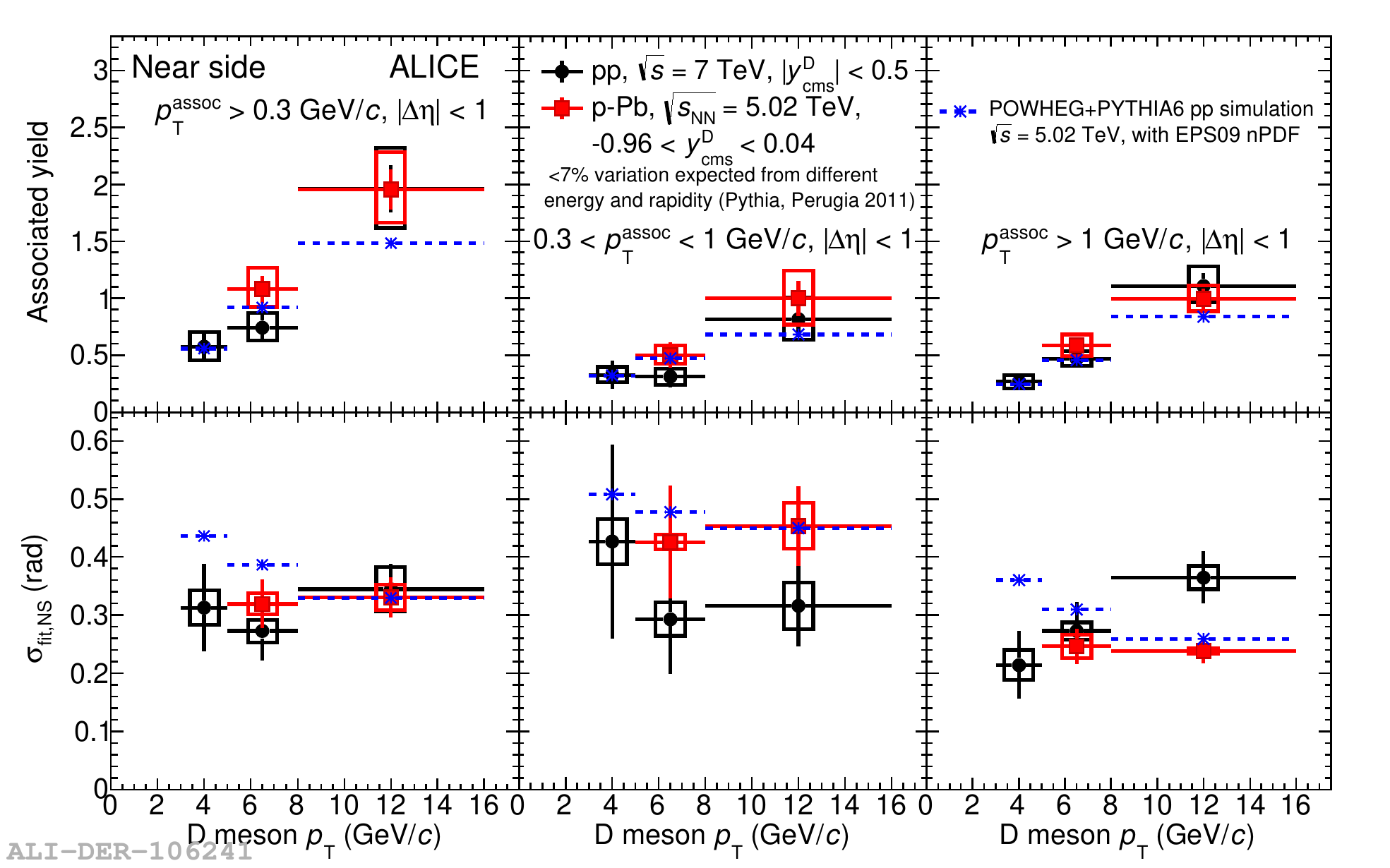}
    \vspace{-0.5em}
    \caption{Comparison of near-side associated yields and widths extracted in pp and $\pPb$ collisions and predicted by PYTHIA+POWHEG simulations as a function of the D-meson $\Pt$ for different associated particle $\Pt$ ranges~\cite{arXivDh}.}
    \label{fig:3}
\end{figure}

Azimuthal correlation distributions of $\Dz$, $\Dp$ and $\Ds$ mesons (trigger particles) with charged particles (associated particles) in $|\deta| <$ 1 are evaluated in pp collisions at $\s$ = 7 TeV and $\pPb$ collisions at $\sNN$ = 5.02 TeV, for different ranges of the D-meson ($3 < \PtD < 16$ $\gevc$ in pp, $5 < \PtD < 16$ $\gevc$ in $\pPb$) and associated particle $\Pt$ (starting from $\PtAssoc > 0.3$ $\gevc$).
The contribution of D-meson combinatorial background is removed by subtracting the correlation distribution evaluated from the sidebands of the D-meson invariant mass distribution. An event-mixing correction is applied to account for detector inhomogeneities and limited acceptance. The distributions are corrected for inefficiencies in the reconstruction and selection of trigger and associated particles.
The contribution of D mesons from beauty-hadron decays is subtracted, exploiting templates of the angular correlations of feed-down D mesons and charged particles obtained from PYTHIA simulations~\cite{PYTH6}, normalized to the expected feed-down contribution evaluated from FONLL calculations~\cite{FONLL}.
The azimuthal correlation distributions of $\Dz$, $\Dp$ and $\Ds$ mesons are compatible within uncertainties. A weighted average of the three D-meson measurements is thus performed to reduce the statistical uncertainty.
The per-trigger azimuthal correlation distributions are fitted with two Gaussian functions and a constant (baseline), allowing us to extract quantitative observables such as the near-side associated yield, near-side peak width and baseline.
Figure~\ref{fig:2} compares the baseline-subtracted D meson-charged particle azimuthal correlation distributions extracted in pp collisions with predictions by PYTHIA6~\cite{PYTH6}, PYTHIA8~\cite{PYTH8} and POWHEG+PYTHIA6~\cite{POWH1,POWH2} simulations, as a function of the D-meson $\Pt$, for different associated particle $\Pt$ ranges~\cite{arXivDh}. The measured correlation pattern is reproduced by the simulations within uncertainties, though a hint for a more pronounced near-side peak in data than in models is visible in the $8 < \PtD < 16$ $\gevc$ range.
A more quantitative comparison of the near-side peak properties can be performed considering the observables extracted from the fit to the correlation distributions. The trends of the near-side associated yields and the near-side peak widths extracted in pp and $\pPb$ collisions versus the D-meson $\Pt$ are compared in Fig.~\ref{fig:3} for different associated particle $\Pt$ ranges. Compatible values of the near-side observables are obtained in pp and $\pPb$ collisions. No modifications of the near-side peaks due to cold nuclear matter effects are observed in $\pPb$ collisions with the current uncertainties.
Predictions by POWHEG+PYTHIA6 simulations, including nuclear shadowing effects for the nucleon parton distribution functions, are also in agreement with the measurements.


\section{Summary}
An extension of the $\Pt$-differential prompt $\Dz$ production cross section measurement down to \mbox{$\Pt$ = 0~$\gevc$} was presented and was found to be compatible with previous ALICE measurements in the common $\Pt$ range. It allowed to reduce the uncertainties on the total charm production cross section.
The D meson-charged particle azimuthal correlation distributions, measured in pp and $\pPb$ collisions, as well as observables describing near-side peak properties extracted from their fits, are in agreement between each other within uncertainties, and are well described by PYTHIA and POWHEG+PYTHIA Monte Carlo simulations.

\end{document}